\documentclass[a4paper]{article}
\pdfoutput=1
\usepackage{INTERSPEECH2019}
\usepackage{color}
\usepackage{balance}

\title{Exploiting Deep Sentential Context for Expressive End-to-End Speech Synthesis}
\name{Fengyu Yang$^1$\thanks{First author performed part of this work at Xiaomi. Lei Xie is the corresponding author. This work was partially supported by the National Key Research and Development Program of China (No.2017YFB1002102).}, Shan Yang$^2$, Qinghua Wu$^3$, Yujun Wang$^3$, Lei Xie$^{1,2}$}
\address{
  Audio, Speech and Language Processing Group (ASLP@NPU),\\
  $^1$School of Software, $^2$School of Computer Science, Northwestern Polytechnical University, China\\
  $^3$Xiaomi, Beijing, China}
\email{\{fyyang, syang, lxie\}@nwpu-aslp.org, \{wuqinghua, wangyujun\}@xiaomi.com}

\begin{document}

\maketitle
\begin{abstract}

 Attention-based seq2seq text-to-speech systems, especially those use self-attention networks (SAN), have achieved state-of-art performance. But an expressive corpus with rich prosody is still challenging to model as 1) prosodic aspects, which span across different sentential granularities and mainly determine acoustic expressiveness, are difficult to quantize and label and 2) the current seq2seq framework extracts prosodic information solely from a text encoder, which is easily collapsed to an averaged expression for expressive contents. In this paper, we propose a context extractor, which is built upon SAN-based text encoder, to sufficiently exploit the sentential context over an expressive corpus for seq2seq-based TTS. Our context extractor first collects prosodic-related sentential context information from different SAN layers and then aggregates them to learn a comprehensive sentence representation to enhance the expressiveness of the final generated speech. Specifically, we investigate two methods of context aggregation: 1) \textit{direct aggregation} which directly concatenates the outputs of different SAN layers, and 2) \textit{weighted aggregation} which uses multi-head attention to automatically learn contributions for different SAN layers. Experiments on two expressive corpora show that our approach can produce more natural speech with much richer prosodic variations, and weighted aggregation is more superior in modeling expressivity.

\end{abstract}
\noindent\textbf{Index\! Terms}: speech synthesis, self-attention network, prosody

\section{Introduction}

Recently, the naturalness of corpus-based text-to-speech (TTS) has been significantly improved with the use of attention-based sequence-to-sequence (seq2seq) mapping framework~\cite{wang2017tacotron,shen2018natural}. Such so-called end-to-end (E2E) systems directly employ a text encoder network to learn linguistic, syntactic and semantic information from simple character or phoneme sequences. The sequence of aggregated textual representation is further attended by an acoustic decoder network through some attention mechanism, producing predicted speech representations (e.g., mel-spectrogram) that are subsequently transformed to waveforms via a neural vocoder.

 \textit{Sentential context}~\cite{wang2019exploiting} mainly involves the latent syntactic and semantic information embedded in the text, recently proved to be important in natural language processing (NLP) tasks~\cite{wang2019exploiting,peters2018deep}. It might be essential to the naturalness of speech synthesis as well, especially for a system built upon an expressive corpus with rich prosodic variations. The seq2seq framework extracts prosodic information solely from the text encoder in an unsupervised way, which is easily collapsed to an averaged expression for expressive contents. To better make use of the sentential context in an E2E framework, one way is feature engineering as the previous generation of TTS does. For example, recent study has shown that exploiting syntactic features in a parsed tree is beneficial to the richness of the prosodic outcomes, leading to more natural synthesized speech~\cite{guo2019exploiting}.


 However, modeling expressiveness in text-to-speech is still challenging as it refers to different levels of syntactic and semantic information reflected in intensity, rhythm, intonation and other prosody related factors. However, it is difficult to define the relations explicitly between the syntactic/semantic factors and the prosodic factors. To model expressivity, the global style tokens (GST) family~\cite{wang2018style,an2019learning} learns style embeddings from a reference audio in an unsupervised way, which lets the synthesized speech imitate the style of reference audio. Although the style embeddings from a reference audio is helpful to control the style of synthesized speech, it is hard to choose an appropriate reference audio for each input sentence. Likewise, the variational autoencoder (VAE) models styles or expressivity in a similar way~\cite{zhang2019learning}.

Recent studies have revealed that \textit{self-attention} based networks (SAN)~\cite{li2019neural,yasuda2019investigation,yang2019enhancing,yang2019improving} have strong ability in capturing global prosodic information, leading to more natural synthesized speech. And unveiled by recent NLP tasks, different SAN encoder layers can capture latent syntactic and semantic properties of the input sentence at different levels~\cite{wang2019exploiting,peters2018deep}. But current SAN-based TTS systems only leverage the highly aggregated latent text representation, usually the outputs of the text encoder, from the simple textual input, to guide the speech generation process. Although the highly aggregated representation can be treated as a global description of the sentential context, it is not enough to generate expressive content according to our experiments as it may disperse the contribution of sentential context embedded in the intermediate SAN layers~\cite{shi2016does}.

In this paper, to excavate the sentential context for expressive speech synthesis, we propose a context extractor to sufficiently exploit sentential context over an expressive corpus for seq2seq-based TTS. Specifically, we utilize different levels of representations from the SAN-based text encoder to build a context extractor, which is helpful to extract different levels of syntactic and semantic information~\cite{dou2018exploiting}. In details, our context extractor first collects the prosodic-related sentential context information from different SAN-based encoder layers, and then aggregates them to learn a comprehensive sentence representation to enhance the expressiveness of the final generated speech. Specifically, we investigate two methods of context aggregation: 1) \textit{direct aggregation} which directly concatenates the outputs of different SAN layers, and 2) \textit{weighted aggregation} which uses multi-head attention to automatically learn contributions for different SAN layers. Experiments on two expressive corpora show that our approach can produce more natural speech with richer prosodic variations, and weighted aggregation is more superior in modeling expressivity. 

\section{Proposed Model}

\begin{figure}[t]
  \centering
  \includegraphics[width=7.5cm]{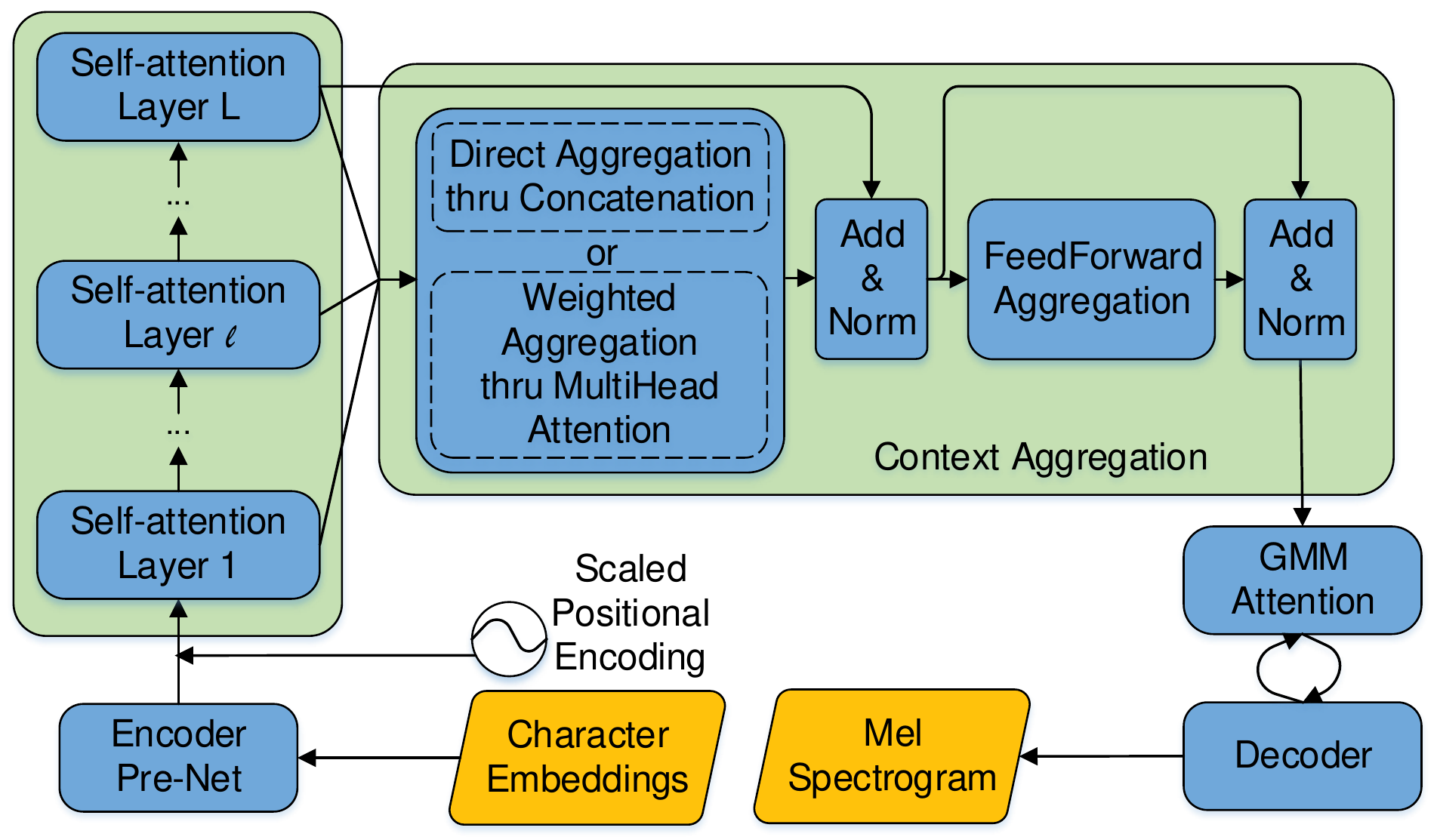}
  \vspace{-0.2cm}
  \caption{Proposed architecture with context aggregation based on Tacotron2 and SAN encoder.}
  \label{fig:SAG-Tacotron2}\vspace{-15pt}
\end{figure}

Figure~\ref{fig:SAG-Tacotron2} illustrates our proposed approach on exploiting deep sentential contexts for expressive speech synthesis. It contains a modified self-attention based text encoder, an auto-regressive decoder and a GMM-based attention~\cite{battenberg2019location} to bridge the encoder and the decoder. WaveGlow~\cite{prenger2019waveglow} is adopted to reconstruct waveforms from mel spectrogram. We augment the encoder with a context aggregation module, which will be described in detail.

\subsection{Self-attention based Encoder}
\label{subsec:self-attention-based-encoder}

Self-attention based sequence-to-sequence framework has been successfully applied to speech synthesis~\cite{li2019neural,yasuda2019investigation,yang2020localness}. In the basic SAN-based text encoder, there is a stack of $L$ blocks, each of which has two sub-networks: a multi-head attention and a feed forward network. The residual connection and layer normalization are applied to both of the sub-networks. Formally, from the previous encoder block output $H^{l-1}$, the first sub-network $C^l$ and the second sub-network $H^l$ are calculated as:

\begin{equation}
  C^l={\rm LN}({\rm MultiHead}(head_1^l,\dots,head_H^l)+H^{l-1}),
  \label{eq:C}
\end{equation}
\begin{equation}
  H^l={\rm LN}({\rm FFN}(C^l)+C^l).
  \label{eq:H}
\end{equation}
where MultiHead($\cdot$), FFN($\cdot$) and LN($\cdot$) are multi-head attention, feed forward network and layer normalization respectively. And each head in multi-head attention split from the previous encoder block output is computed by:
\begin{equation}
  head_h={\alpha}{\cdot}V={\rm softmax}(\frac{QK^T}{\sqrt{d}}){\cdot}V,
  \label{eq:head}
\end{equation}
where $\{Q,K,V\}$ represent queries, keys and values, $d$ is the dimension of the hidden state and ${\alpha}$ represents the weight matrix for each head.

\subsection{Direct Aggregation}
\label{subsec:context-extraction-module}

Although the SANs have the ability of directly capturing global dependencies among whole input sequence~\cite{vaswani2017attention}, it may not appropriately exploit the sentential context because it calculates the relevance between the characters without considering the contextual information~\cite{wang2019exploiting,dou2018exploiting}. Besides, the weighted sum option from the lower layers in SANs has only aggregated the global contextual information, which may weaken the contribution of sequential context extracted in each block.

To fully make use of the contexts extracted from each block, we propose a context extractor to aggregate the different levels of contexts to form a comprehensive sentence representation. For the $l^{th}$ self-attention block, we extract the intermediate context from the output $H^l$ through:
\begin{equation}
  g^l={\rm g}(H^l)={\rm MeanPool}({\rm Conv1d}(H^l)),
  \label{eq:gl}
\end{equation}
where Conv1d means 1d-convolution, MeanPool represents mean pooling\cite{iyyer2015deep}, g($\cdot$) denotes the function to summarize the outputs of self-attention layers, and $g^l$ represents the sentential context from $l^{th}$ block. A straightforward and intuitive choice to aggregate the different levels of contexts is through a concatenation operation, with residual connection and layer normalization~\cite{li2019neural}:
\begin{equation}
  C_g={\rm LN}({\rm Concat}( g^0,\dots,g^L)+g^L),
  \label{eq:concat}
\end{equation}
where $g^0$ represents the inputs of the first self-attention layer through Eq.~(\ref{eq:gl}).
To further integrate the information concatenated from all sentential contexts, we use another round of feed-forward network and layer normalization as the final aggregation function~\cite{dou2018exploiting}\cite{wang2020multi}:
\begin{equation}
  g={\rm LN}({\rm FFN}(C_g)+C_g).
  \label{eq:deep-ffn}
\end{equation}
Here, $g$ is the final sentential context.



\subsection{Weighted Aggregation}
\label{subsec:deep-sentential-context}

With direct aggregation, the sequential contexts of each block are simply concatenated to guide the auto-regressive generation, which does not consider the varying importance of each $g^l$. Assuming the sequential contexts in each block may have different contribution to the expressiveness of the synthesized speech, we utilize a self-learned weighted aggregation module across layers to catch the different levels of contribution.

In detail, we employ a multi-head attention to learn the contribution of each block. The individual sentential contexts $\{g^0,g^1,\dots,g^L\}$ are treated as attention memory for the attention based weighted aggregation. Specifically, we transpose the dimension of sequential length with the number of heads in the multi-head attention to combine the contextual information across layers. Therefore, we modify Eq.~(\ref{eq:concat}) to obtain the weighted contexts:
\begin{equation}
  C_g={\rm LN}({\rm MultiHead}( g^0,\dots,g^L)+g^L),
  \label{eq:weighted}
\end{equation}
where the modified $C_g$ offers a more precise control of aggregation for each $g^l$.

\section{Experiments}

\subsection{Basic setups}

To investigate the effectiveness of modeling expressiveness, we carried out experiments on two expressive Mandarin corpora -- the publicly-available Blizzard Challenge 2019 corpus~\cite{wu2019blizzard} from a male talk-show speaker and an internal voice assistant corpus from a female speaker. The talk-show (TS) corpus contains about 8 hours speech of, and the voice assistant (VA) corpus contains about 40 hours of speech. Both corpora are expressive in prosodic delivery, separated to non-overlapping training and testing sets (with data ratio 9:1). Besides, we also utilize a publicly-available standard reading-style corpus~\cite{DB1} with less expressivity to see how our approach perform. The corpus, named DB1, contains 10 hours of female speech with consistent reading style. Again, we separate the corpus to training and testing with ratio 9:1. Linguistic inputs include phones, tones, character segments and three levels of prosodic segments: prosodic word (PW), phonological phrase (PPH) and intonation phrase (IPH). We use 80-band mel-spectrogram extracted from 22.05KHz waveforms (for TS and VA) and 16KHz waveforms (for DB1) as acoustic targets. We calculate mel cepstral distortion (MCD) on test set for objective evaluation. As for subjective evaluation, we conduct mean opinion score (MOS) and A/B preference test on 30 randomly selected  test set samples and 20 native Chinese listeners participated in the tests.
\subsection{Model details}

We use the standard encoder-decoder structure in Tacotron2~\cite{shen2018natural} as the baseline, but GMM attention is adopted instead because it can bring superior naturalness and stability~\cite{battenberg2019location}. For networks using SAN based encoder, a 3-layer CNN is firstly applied to the input text embeddings with positional information. Each self-attention block includes an 8-head self-attention and a feed forward sub-network consisting of two linear transformations with 2048 and 512 hidden units. Residual connection and layer normalization are applied to these two sub-networks.  There are totally 6 self-attention blocks. In the aggregation module, we double feed $g^L$ into aggregation attention function for the convenience of implementation, where the number of heads in multi-head attention are $length$ and the dimension of weighted matrix are $[batch, length, 8, 8]$. For the remaining part, we adopt the auto-regressive decoder described in~\cite{shen2018natural}. We use WaveGlow as vocoder which follows the structure in~\cite{prenger2019waveglow}, trained using the same training set. We built the following systems for comparison:

\begin{itemize}
\item  {\bf Base}: Baseline system following Tacotron2~\cite{shen2018natural} with slightly modified GMMv2 attention~\cite{battenberg2019location}.
\item  {\bf SA}: Another baseline system with SAN based encoder described in Section~\ref{subsec:self-attention-based-encoder}.
\item  {\bf SA-DA}: SAN based encoder with the direct aggregation module fusing all sentential contexts described in Section~\ref{subsec:context-extraction-module}.
\item  {\bf SA-WA}: SAN based encoder with the weighted aggregation module fusing all sentential contexts described in Section~\ref{subsec:deep-sentential-context}.
\end{itemize}

\subsection{Objective Evaluation}
\begin{table}[t]
  \caption{MCD scores over the two expressive corpora.}
  \vspace{-0.2cm}
  \label{tab:mcd}
  \centering
  \begin{tabular}{c|c|c|c|c}
  \hline
     Corpus & BASE & SA & SA-DA & SA-WA \\\hline
    TS & 8.01 & 7.48 & 7.42 & \textbf{7.32} \\ 
    VA & 7.60 & 7.37 & 7.32 & \textbf{7.23} \\\hline 
  \end{tabular}
  \vspace{-15pt}
\end{table}
Table~\ref{tab:mcd} shows the MCD results of different systems. It demonstrates that SAN based encoder has lower MCD than the RNN based encoder for both expressive corpora. It also shows that modeling sentential context can further improve the performance of SAN based encoder. Besides, weighted aggregation is a better way than direct aggregation to extract the deep sentential context. With the help of deep sentential context, the SA-WA system achieves the lowest MCD on both expressive corpora, which indicates that the synthesized speech samples are the most similar ones to the real speech samples.

\subsection{Subjective Evaluation}

\begin{figure}[t]
  \centering
  \includegraphics[width=6.5cm]{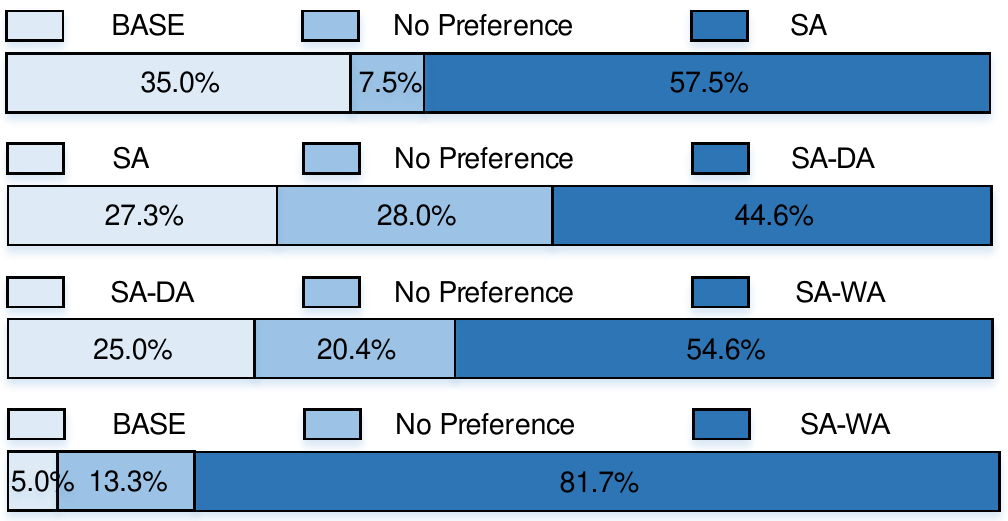}
  \vspace{-0.2cm}
  \caption{AB Preference results on TS with confidence intervals of 95\% and p-value $\textless$ 0.0001 from t-test.}
  \vspace{-10pt}
  \label{fig:ABTest-l}
  \vspace{-0.15cm}
\end{figure}

\begin{figure}[t]
\vspace{-0.05cm}
  \centering
  \includegraphics[width=6.5cm]{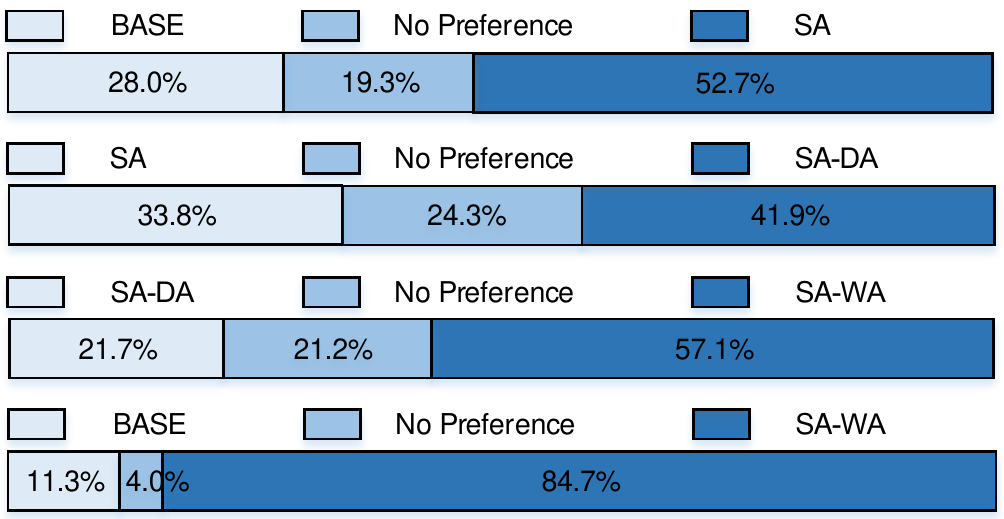}
  \vspace{-0.2cm}
  \caption{AB Preference results on VA with confidence intervals of 95\% and p-value $\textless$ 0.0001 from t-test.}
  \label{fig:ABTest-x}\vspace{-8pt}
\end{figure}

We conduct AB preference tests and MOS tests on the two expressive test sets which include a large number of modal particles, interrogatives and exclamations. The listeners are asked to select preferred audio according to the overall impression on the expressiveness of the testing samples\footnote{Samples can be found from: https://fyyang1996.github.io/context/}. The AB preference results are shown in Figure~\ref{fig:ABTest-l} and~\ref{fig:ABTest-x} for TS and VA, respectively. MOS results are reported in Table~\ref{tab:mos}.

\begin{table}[t]
  \caption{The MOS over the two expressive corpora with confidence intervals of 95\%.}
  \vspace{-0.2cm}
  \label{tab:mos}
  \centering
  \begin{tabular}{c|c|c|c}
  \hline
    Corpus & BASE & SA & SA-DA \\\hline
    TS & 3.84$\pm$0.05 & 3.97$\pm$0.06 & 4.04$\pm$0.06 \\
    VA & 4.11$\pm$0.06 & 4.20$\pm$0.06 & 4.24$\pm$0.06 \\\hline
    Corpus & SA-WA & GT \\\hline
    TS & \textbf{4.11$\pm$0.06} & 4.54$\pm$0.05 \\
    VA & \textbf{4.36$\pm$0.07} & 4.63$\pm$0.05 \\\hline
  \end{tabular}
  \vspace{-0.2cm}
\end{table}

For baseline systems, we can find that the SA system with SAN based encoder brings better performance on expressiveness than the conventional BASE system. It indicates that using self-attention layers as text encoder may capture features that better represents expressiveness, in accordance with our previous findings~\cite{yang2019improving}. For the proposed context extractor, we find that, by introducing direct aggregation across all the self-attention layers, system SA-DA achieves significantly better performance than the solely self-attention based encoder system SA. This is confirmed by both AB preference and MOS test on the two tested corpora. By further replacing simple concatenation operation with multi-head attention aggregation (i.e., weighted aggregation), system SA-WA brings extra performance gain over system SA-DA. Listeners particularly give the SA-WA system more preferences according to the AB preference test. In summary, the results unveil that the deep sentential context encoder achieves significantly better performance than the baseline systems, showing that modeling different levels of latent syntax and semantic information through a deep encoder is effective for generating expressive speech. This conclusion is consistently confirmed on two expressive corpora.

\subsection{Performance on less-expressive corpus}

\begin{figure}[t]
  \centering
  \includegraphics[width=6.5cm]{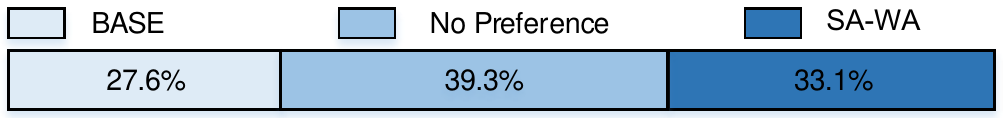}
  \vspace{-0.2cm}
  \caption{AB Preference results on DB1 with confidence intervals of 95\% and p-value $\textless$ 0.0001 from t-test.}
  \label{fig:ABTest-d}\vspace{-15pt}
  \vspace{-0.1cm}
\end{figure}

We also quickly examine the performance of our approach on a less-expressive reading-style corpus -- DB1~\cite{DB1}, to see how our sentential context extractor perform. Here, we only compare the above best-performed SA-WA system with the BASE system. The MCD scores for BASE and SA-WA are 5.78 and 5.72, respectively. The AB preference is illustrated in Figure~\ref{fig:ABTest-d}. Interestingly, the effectiveness of our sentential context extractor is not salient on this less-expressive corpus, which is proved by close MCD and AB preference scores between the two systems. In other words, our sentential context extractor works better on expressive datasets, which is further confirmed in the following.

\subsection{Analysis}

\textbf{Prosody Correlation } To further evaluate the expressiveness for statistical significance, we extract the acoustic features commonly associated with prosody: relative energy within each phoneme (E), duration in ms (Dur.) and fundamental frequency in Hertz (F0), which represent phoneme-level intensity, rhythm and intonation of audio, respectively. According to \cite{sun2020generating}, we measure the three prosody attributes for each phoneme throughout additional alignments. The ratio of the average signal magnitude within a phoneme with the average magnitude of the entire utterance is used as the relative energy of a phoneme. We calculate the number of frames within a phoneme as the duration of the phoneme. And the mean value of F0 within a phoneme is regarded as a prosody attribute. To estimate these statistics, we synthesized 100 random samples in the test set to calculate the Pearson correlation coefficient between all systems and the ground truth. The higher Pearson correlation coefficient value the model produces, the higher accuracy of the predicted prosody attribute the model can achieve.

\begin{table}[t]
  \caption{Correlation in relative energy, duration and F0 within a phoneme computed from different models on TS.}
  \vspace{-0.2cm}
  \label{tab:correlation-lzy}
  \centering
  \begin{tabular}{c|c|c|c|c}
  \hline
     & BASE & SA & SA-DA & SA-WA \\\hline
    E & 0.755 & 0.776 & 0.781 & \textbf{0.799} \\ 
    Dur. & 0.617 & 0.638 & 0.641 & \textbf{0.654} \\ 
    F0 & 0.42 & 0.426 & 0.437 & \textbf{0.501} \\\hline 
  \end{tabular}
  \vspace{-5pt}
\end{table}

Table~\ref{tab:correlation-lzy} shows that our proposed SA-WA system obtains highest correlation scores in all three prosody attributes, which demonstrates that our approach has the best reconstruction performance in phoneme-level intensity, rhythm and intonation. Additionally, \cite{sun2020fully} reveals that the order of prosody attributes being captured is always found to be energy, duration, and F0. Energy is the amplitude of the signal directly related to the reconstruction loss and is easier to be captured, but F0 is the most difficult to be captured as it is modeled implicitly. However, our SA-WA system achieves approximately 20\% gains than the BASE system in F0, which is far more than the promotion of approximately 6\% in energy and duration. Based on these, we believe that the proposed approach has strong ability in modeling F0 that is most difficult one to be captured in the three prosody attributes.

\begin{figure}[t]
  \centering
  \includegraphics[width=7cm,height=4cm]{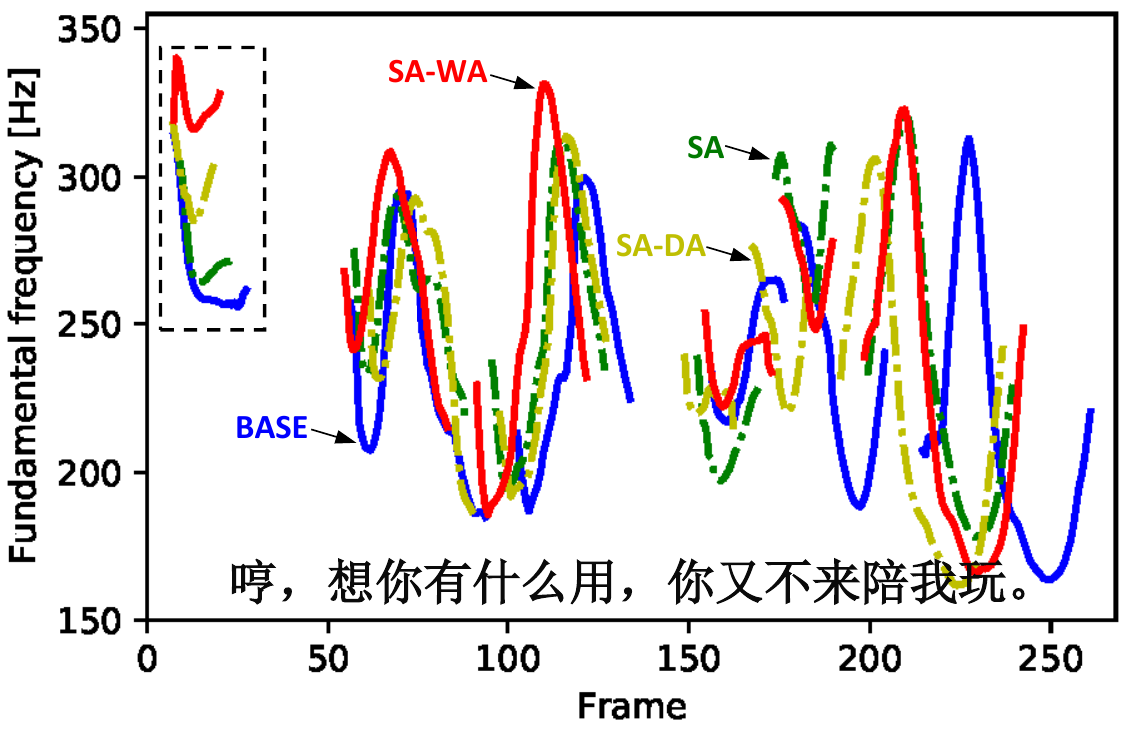}\vspace{-5pt}
  \vspace{-0.2cm}
  \caption{F0 values of a test utterance generated by different systems. Audios can be found in Section 1.1 of the demo page.}\vspace{-5pt}
  \label{fig:pitch}\vspace{-10pt}
\end{figure}

Figure~\ref{fig:pitch} shows the F0 trajectories for a synthesized test utterance. The sentence begins with a modal particle (heng1), which reveals sense of disgust emotion in Mandarin. In this case, high raise of F0, where the SA-WA system produces, can deliver more disgust mood to listeners. This example shows that the proposed sequential context extractor can model better expressive patterns compared to the baseline systems.

\textbf{Prosody Diversity} An expressive TTS system should be able to generate speech with a large prosody diversity. Consequently, we measure the standard deviation of three prosody attributes at phoneme level as well according to \cite{sun2020generating}. And the average standard deviation across all 100 utterances is reported for statistical significance. Table~\ref{tab:diversity-lzy} demonstrate that the SA-WA system has the highest diversity in phoneme-level intensity, rhythm and intonation among all systems, which is the closest to the ground-truth (GT). We believe that the SA-WA system has better ability in modeling prosody variations on expressive datasets. \vspace{-5pt}

\begin{table}[h]
  \caption{Diversity values using average standard deviation computed across 100 samples on TS.}
  \label{tab:diversity-lzy}\vspace{-5pt}
  \centering
  \begin{tabular}{c|c|c|c|c|c}
    \hline
     & BASE & SA & SA-DA & SA-WA & GT \\\hline
    E & 0.238 & 0.277 &  0.285 & \textbf{0.304} & 0.321 \\ 
    Dur. & 33.374 & 34.337 & 34.955 & \textbf{37.003} & 41.866 \\ 
    F0 & 32.362 & 33.405 & 35.161 & \textbf{35.766} & 36.824 \\\hline 
  \end{tabular}
  \vspace{-15pt}
\end{table}

\section{Conclusion}

Seq2seq-based TTS directly maps the character/phoneme sequence to the acoustic feature sequence using an encoder-decoder paradigm. The encoder functions as a sentential context extractor which aggregates latent semantic and syntactic information that highly correlates with the expressiveness of the synthesized speech by the decoder. In this paper, we propose a context extractor, which is built upon the SAN-based text encoder, to sufficiently exploit the text-side sentential context to produce more expressive speech. With the belief that different self-attention layers may capture different levels of latent syntactic and semantic information, which was discovered by recent NLP researches, we proposed two context aggregation strategies: 1) \textit{direct aggregation} which directly concatenates the outputs of different SAN layers, and 2) \textit{weighted aggregation} which uses multi-head attention to automatically learn contributions for different SAN layers. Experiments on two expressive corpora show that the two strategies can produce more natural and expressive speech, and weighted aggregation is more superior. Comprehensive analysis on the synthesized speech demonstrates that our sentential context extractor has better ability in reconstruction of prosody related acoustic features and modeling prosody diversity.

\bibliographystyle{IEEEtran}
\balance
\bibliography{mybib}

\begin{thebibliography}{10}
\providecommand{\url}[1]{#1}
\csname url@samestyle\endcsname
\providecommand{\newblock}{\relax}
\providecommand{\bibinfo}[2]{#2}
\providecommand{\BIBentrySTDinterwordspacing}{\spaceskip=0pt\relax}
\providecommand{\BIBentryALTinterwordstretchfactor}{4}
\providecommand{\BIBentryALTinterwordspacing}{\spaceskip=\fontdimen2\font plus
\BIBentryALTinterwordstretchfactor\fontdimen3\font minus
  \fontdimen4\font\relax}
\providecommand{\BIBforeignlanguage}[2]{{%
\expandafter\ifx\csname l@#1\endcsname\relax
\typeout{** WARNING: IEEEtran.bst: No hyphenation pattern has been}%
\typeout{** loaded for the language `#1'. Using the pattern for}%
\typeout{** the default language instead.}%
\else
\language=\csname l@#1\endcsname
\fi
#2}}
\providecommand{\BIBdecl}{\relax}
\BIBdecl

\bibitem{wang2017tacotron}
Y.~Wang, R.~Skerry-Ryan, D.~Stanton, Y.~Wu, R.~J. Weiss, N.~Jaitly, Z.~Yang,
  Y.~Xiao, Z.~Chen, S.~Bengio \emph{et~al.}, ``Tacotron: Towards end-to-end
  speech synthesis,'' in \emph{2017 International Speech Communication
  Association (Interspeech)}.\hskip 1em plus 0.5em minus 0.4em\relax ISCA,
  2017, pp. 4006--4010.

\bibitem{shen2018natural}
J.~Shen, R.~Pang, R.~J. Weiss, M.~Schuster, N.~Jaitly, Z.~Yang, Z.~Chen,
  Y.~Zhang, Y.~Wang, R.~Skerrv-Ryan \emph{et~al.}, ``Natural tts synthesis by
  conditioning wavenet on mel spectrogram predictions,'' in \emph{2018 IEEE
  International Conference on Acoustics, Speech and Signal Processing
  (ICASSP)}.\hskip 1em plus 0.5em minus 0.4em\relax IEEE, 2018, pp. 4779--4783.

\bibitem{wang2019exploiting}
X.~Wang, Z.~Tu, L.~Wang, and S.~Shi, ``Exploiting sentential context for neural
  machine translation,'' in \emph{2019 Association for Computational
  Linguistics (ACL): Long Papers}.\hskip 1em plus 0.5em minus 0.4em\relax ACL,
  2019, pp. 6197--6203.

\bibitem{peters2018deep}
M.~E. Peters, M.~Neumann, M.~Iyyer, M.~Gardner, C.~Clark, K.~Lee, and
  L.~Zettlemoyer, ``Deep contextualized word representations,'' vol.~1, pp.
  2227--2237, 2018.

\bibitem{guo2019exploiting}
H.~Guo, F.~K. Soong, L.~He, and L.~Xie, ``Exploiting syntactic features in a
  parsed tree to improve end-to-end tts,'' in \emph{2019 International Speech
  Communication Association (Interspeech)}.\hskip 1em plus 0.5em minus
  0.4em\relax ISCA, 2019, pp. 4460--4464.

\bibitem{wang2018style}
Y.~Wang, D.~Stanton, Y.~Zhang, R.~Skerry-Ryan, E.~Battenberg, J.~Shor, Y.~Xiao,
  F.~Ren, Y.~Jia, and R.~A. Saurous, ``Style tokens: Unsupervised style
  modeling, control and transfer in end-to-end speech synthesis,'' in
  \emph{2018 International Conference on Machine Learning (ICML)},
  vol.~80.\hskip 1em plus 0.5em minus 0.4em\relax PMLR, 2018, pp. 5167--5176.

\bibitem{an2019learning}
X.~An, Y.~Wang, S.~Yang, Z.~Ma, and L.~Xie, ``Learning hierarchical
  representations for expressive speaking style in end-to-end speech
  synthesis,'' in \emph{2019 IEEE Automatic Speech Recognition and
  Understanding Workshop (ASRU)}.\hskip 1em plus 0.5em minus 0.4em\relax IEEE,
  2019, pp. 184--191.

\bibitem{zhang2019learning}
Y.-J. Zhang, S.~Pan, L.~He, and Z.-H. Ling, ``Learning latent representations
  for style control and transfer in end-to-end speech synthesis,'' in
  \emph{2019 IEEE International Conference on Acoustics, Speech and Signal
  Processing (ICASSP)}.\hskip 1em plus 0.5em minus 0.4em\relax IEEE, 2019, pp.
  6945--6949.

\bibitem{li2019neural}
N.~Li, S.~Liu, Y.~Liu, S.~Zhao, and M.~Liu, ``Neural speech synthesis with
  transformer network,'' in \emph{2019 The Thirty-Third Conference on
  Artificial Intelligence (AAAI)}, vol.~33.\hskip 1em plus 0.5em minus
  0.4em\relax AAAI, 2019, pp. 6706--6713.

\bibitem{yasuda2019investigation}
Y.~Yasuda, X.~Wang, S.~Takaki, and J.~Yamagishi, ``Investigation of enhanced
  tacotron text-to-speech synthesis systems with self-attention for pitch
  accent language,'' in \emph{2019 IEEE International Conference on Acoustics,
  Speech and Signal Processing (ICASSP)}.\hskip 1em plus 0.5em minus
  0.4em\relax IEEE, 2019, pp. 6905--6909.

\bibitem{yang2019enhancing}
S.~Yang, H.~Lu, S.~Kang, L.~Xie, and D.~Yu, ``Enhancing hybrid self-attention
  structure with relative-position-aware bias for speech synthesis,'' in
  \emph{2019 IEEE International Conference on Acoustics, Speech and Signal
  Processing (ICASSP)}.\hskip 1em plus 0.5em minus 0.4em\relax IEEE, 2019, pp.
  6910--6914.

\bibitem{yang2019improving}
F.~Yang, S.~Yang, P.~Zhu, P.~Yan, and L.~Xie, ``Improving mandarin end-to-end
  speech synthesis by self-attention and learnable gaussian bias,'' in
  \emph{2019 IEEE Automatic Speech Recognition and Understanding Workshop
  (ASRU)}.\hskip 1em plus 0.5em minus 0.4em\relax IEEE, 2019, pp. 208--213.

\bibitem{shi2016does}
X.~Shi, I.~Padhi, and K.~Knight, ``Does string-based neural mt learn source
  syntax?'' in \emph{2016 Empirical Methods in Natural Language Processing
  (EMNLP)}.\hskip 1em plus 0.5em minus 0.4em\relax ACL, 2016, pp. 1526--1534.

\bibitem{dou2018exploiting}
Z.-Y. Dou, Z.~Tu, X.~Wang, S.~Shi, and T.~Zhang, ``Exploiting deep
  representations for neural machine translation,'' in \emph{2018 Empirical
  Methods in Natural Language Processing (EMNLP)}.\hskip 1em plus 0.5em minus
  0.4em\relax ACL, 2018, pp. 4253--4262.

\bibitem{battenberg2019location}
E.~Battenberg, R.~Skerry-Ryan, S.~Mariooryad, D.~Stanton, D.~Kao, M.~Shannon,
  and T.~Bagby, ``Location-relative attention mechanisms for robust long-form
  speech synthesis,'' \emph{arXiv preprint arXiv:1910.10288}, 2019.

\bibitem{prenger2019waveglow}
R.~Prenger, R.~Valle, and B.~Catanzaro, ``Waveglow: A flow-based generative
  network for speech synthesis,'' in \emph{2019 IEEE International Conference
  on Acoustics, Speech and Signal Processing (ICASSP)}.\hskip 1em plus 0.5em
  minus 0.4em\relax IEEE, 2019, pp. 3617--3621.

\bibitem{yang2020localness}
S.~Yang, H.~Lu, S.~Kang, L.~Xue, J.~Xiao, D.~Su, L.~Xie, and D.~Yu, ``On the
  localness modeling for the self-attention based end-to-end speech
  synthesis,'' \emph{Neural Networks}, vol. 125, pp. 121--130, 2020.

\bibitem{vaswani2017attention}
A.~Vaswani, N.~Shazeer, N.~Parmar, J.~Uszkoreit, L.~Jones, A.~N. Gomez,
  {\L}.~Kaiser, and I.~Polosukhin, ``Attention is all you need,'' in \emph{2017
  Advances in neural information processing systems (NIPS)}, 2017, pp.
  5998--6008.

\bibitem{iyyer2015deep}
M.~Iyyer, V.~Manjunatha, J.~Boyd-Graber, and H.~Daum{\'e}~III, ``Deep unordered
  composition rivals syntactic methods for text classification,'' in \emph{2015
  the 53rd Annual Meeting of the Association for Computational Linguistics and
  the 7th International Joint Conference on Natural Language Processing (ACL):
  Long Papers}, vol.~1.\hskip 1em plus 0.5em minus 0.4em\relax ACL, 2015, pp.
  1681--1691.

\bibitem{wang2020multi}
Q.~Wang, F.~Li, T.~Xiao, Y.~Li, Y.~Li, and J.~Zhu, ``Multi-layer representation
  fusion for neural machine translation,'' \emph{arXiv preprint
  arXiv:2002.06714}, 2020.

\bibitem{wu2019blizzard}
Z.~Wu, S.~Le~Maguer, J.~Cabral, and S.~King, ``The blizzard challenge 2019,''
  2019.

\bibitem{DB1}
D.~Baker, ``The db1 chinese female speech dataset,''
  \url{https://www.data-baker.com/us.html}, 2020.

\bibitem{sun2020generating}
G.~Sun, Y.~Zhang, R.~J. Weiss, Y.~Cao, H.~Zen, A.~Rosenberg, B.~Ramabhadran,
  and Y.~Wu, ``Generating diverse and natural text-to-speech samples using a
  quantized fine-grained vae and auto-regressive prosody prior,'' \emph{arXiv
  preprint arXiv:2002.03788}, 2020.

\bibitem{sun2020fully}
G.~Sun, Y.~Zhang, R.~J. Weiss, Y.~Cao, H.~Zen, and Y.~Wu, ``Fully-hierarchical
  fine-grained prosody modeling for interpretable speech synthesis,''
  \emph{arXiv preprint arXiv:2002.03785}, 2020.

\end{thebibliography}


\end{document}